\documentclass[final,authoryear,3p,times,twocolumn]{elsarticle}

\usepackage{graphics}

\usepackage{amssymb}
\usepackage{amsthm}
\usepackage{wasysym}

\journal{Geochimica et Cosmochimica Acta}

\begin{document}

\begin{frontmatter}


\title{Prediction of equilibrium Li isotope fractionation between minerals and aqueous solutions at high $P$ and $T$: an efficient {\it ab initio} approach}



\author{Piotr M. Kowalski and Sandro Jahn}

\address{GFZ German Research Centre for Geosciences, Telegrafenberg, 14473 Potsdam, Germany}

\begin{abstract}
The mass-dependent equilibrium stable isotope fractionation between different materials is an important geochemical process.
Here we present an efficient method to compute the 
isotope fractionation between complex minerals and fluids at high pressure, $P$, and temperature, $T$, representative for the Earth's crust and mantle.
The method is tested by computation of the equilibrium fractionation of lithium isotopes
between aqueous fluids and various Li bearing minerals such as staurolite, spodumene and mica. We are able to correctly predict 
the direction of the isotope fractionation as observed in the experiments.
On the quantitative level the computed fractionation factors agree within $1.0\,\permil$ with 
the experimental values indicating predictive power of {\it ab initio} methods. We show that with {\it ab initio} methods 
we are able to investigate the underlying mechanisms driving the equilibrium isotope
fractionation process, such as coordination of the fractionating elements, their bond strengths to the neighboring atoms, compression of fluids 
and thermal expansion of solids.
This gives valuable insight into the processes governing the isotope fractionation mechanisms on the atomic scale.
The method is applicable to any state and does not require different treatment of crystals and fluids.

\end{abstract}

\begin{keyword}


\end{keyword}

\end{frontmatter}


\section{Introduction}
\label{}
The fractionation of stable isotopes between various materials is of importance in geoscience,
as the variation in isotope content provides valuable information on processes and interaction between atmosphere, 
biosphere, geosphere and hydrosphere. Although there is substantial analytical work performed in this area, 
reliable computational methods to predict isotope fractionation factors have been available only recently,
proving that they can contribute towards understanding geochemical mechanisms responsible for production 
of isotope signatures \citep{D97,YMMW01,S04,DK08,HS08,ML09,SMH09,Z09,HS10,RC10,RB10,Z09,Z10}.

\begin{table*}[t]
\caption{Lattice parameters of the investigated Li-bearing silicates. 
$N_{atoms}$ is the number of atoms in the modeled supercell. The units are $\rm \AA$ and degrees.}
\label{T1}      
\centering          
\begin{tabular}{lcccccc}     
\hline\hline       
& staurolite & spodumene & mica 1M & mica 2M1 & mica 2M2 & mica 3T \\
\hline                    
 a &  7.848 &  9.463  & 5.20  & 5.209  & 9.04  & 5.200  \\
 b & 16.580 &  8.392  & 9.01  & 9.053  & 5.22  & 5.200  \\
 c &  5.641 & 10.436  & 10.09  & 20.053  & 20.2100  & 29.760  \\
 $\alpha$ & 90  & 90   & 90  & 90  & 90  & 90  \\
 $\beta$  & 90  & 110.15  & 99.38  & 95.74  & 99.58  & 90  \\
 $\gamma$ & 90  & 90   & 90  & 90  & 90  & 120  \\
 ref.    & $^1$  & $^2$ & $^3$ & $^3$ & $^4$ & $^5$ \\
 $N_{atoms}$ & 81& 80& 44& 88& 88& 66\\
\hline                  
\end{tabular}
\\
References: $^1$\cite{CM02}, $^2$\cite{CS73},$^3$\cite{S76},$^4$\cite{SF73},$^5$\cite{B78}
\end{table*}

{\it Ab initio} calculations of equilibrium isotope fractionation between minerals have received considerable attention recently.
Previous studies, however, were mostly limited to simple crystals containing just a few atoms in the unit cell 
such as quartz, kaolinite or carbonate minerals \citep{ML07,RC10}, as the methods used require
considerable computational resources. Only very recently, the calculations have been extended to more complex crystalline solids
containing up to 80 atoms in the unit cell by \citet{SCh11}. 
There are different approaches used in the computation of the mass-dependent stable isotope equilibrium fractionation factors of minerals,
but all methods require knowledge of the vibrational spectrum of the considered system, which is usually computed using {\it ab initio} methods.
\citet{ML07} performed full 
normal mode analysis of the solid phases accounting for the phonon dispersion in reciprocal space.
Because of the huge computational requirements, this method, although correct, can be only applied to the computation of stable isotope fractionation between simple phases. 
On the other hand, in order to
derive the frequencies required for the computation of the fractionation factors \citet{RC10} 
approximated the solids by small clusters and treated them as large molecules. This approach is based on well established 
theories of stable isotope fractionation \citep{BM47,K82,CCH01} showing that the major contribution to the mass-dependent 
fractionation comes from the local vibrational motion of the fractionating element. In line with this finding \citet{SCh11} has found that considering
the phonon spectrum on a single phonon wave vector only
is sufficient for modeling of $^{26}$Mg/$^{24}$Mg isotope fractionation between magnesium bearing crystal phases.

\begin{figure*}[t]
\includegraphics[angle=270,width=4.in]{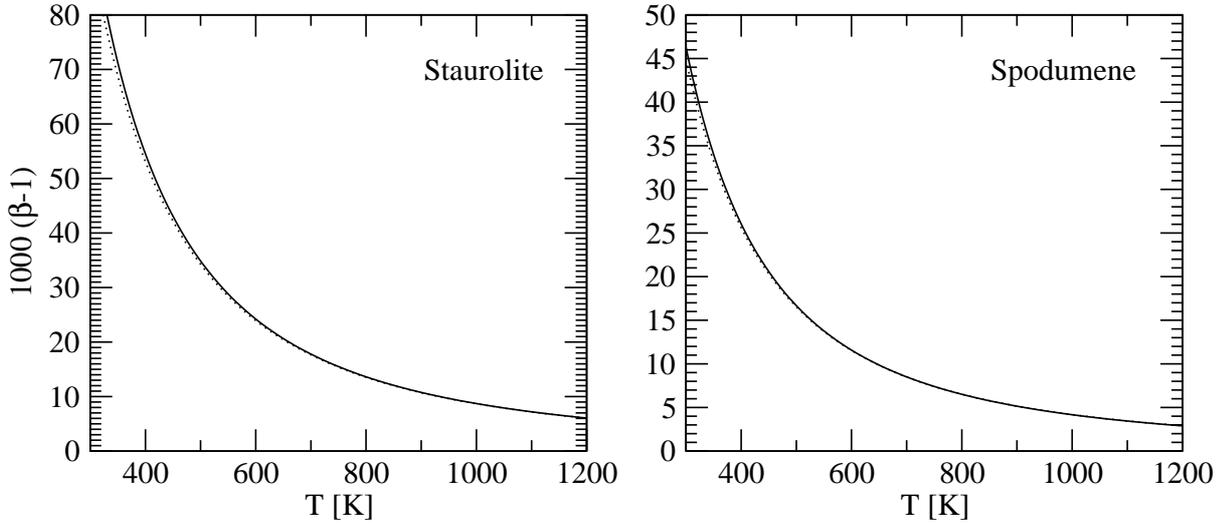}
\caption{ The $\beta$ factors for staurolite and spodumene. The lines represent the results 
using Eq. \ref{BAPX} (solid lines) and Eq. \ref{beta} with the full frequency spectrum (dotted lines). \label{FIG1}}
\end{figure*}

As fluid-rock interactions are a major cause that alter the isotopic signature of a mineral in a rock,
understanding the equilibrium isotope fractionation processes between minerals and aqueous fluids is of great importance in petrology.
Although there has been considerable work on stable isotope fractionation between
various minerals and the computational techniques are well established, the question of treating fluids, namely aqueous solutions remains open.
Most of the {\it ab initio} calculations of isotope fractionation in fluids 
use the cluster approach \citep{DK08,HS08,Z09,RB10,RC10,HS10,YMMW01,Z10}, in which the considered species 
(ions or molecular complexes such as Fe, Mg, $\rm H_3BO_3$) are surrounded by a hydration shell and the 
whole structure is relaxed assuming $T=0$~K. 
This approach is based on the computation of static atomic configurations and is valid at low temperatures only.
In case of Li in aqueous solution at high temperatures ($T\sim1000\rm\,K$), frequent exchange between particles of the hydration shell 
surrounding Li cation with the fluid is observed on time scales as short as picoseconds ($10^{-12}$~s, \citet{JW09}).
Distribution of cation coordination and cation-O bond lengths, effects that are expected to affect the isotope fractionation \citep{BM47}, 
also change with pressure \citep{JW09,WJ11}. These features are difficult to account for by using the cluster approximation for a compressible fluid at high temperature.
The impact of the dynamical behavior of particles and compressibility of fluid must be investigated
in order to properly compute the isotope fractionation in aqueous fluids.
The only recent {\it ab initio} 
work that accounts for the dynamical effects on the isotope fractionation in fluid is by \citet{RB07}
who considered boron isotope fractionation between $\rm B(OH)_3$ and $\rm B(OH)_4^-$ in aqueous solution.
They performed {\it ab initio} molecular dynamics simulations of this system and tried to use the vibrational density of states 
derived through the Fourier transform of the velocity auto-correlation function as an input for the calculation of the 
$^{11}$B/$^{10}$B isotope fractionation coefficient. The resulting fractionation factor $\alpha=0.86$  
is much lower than the experimental value
$\alpha=1.028$. Interestingly, the discrepancy between experiment and theory is cured by quenching the selected configurations
along the molecular dynamics trajectory and 
computing the harmonic frequencies. The fractionation factor derived using these frequencies 
exactly reproduces the experimental value.

In this contribution we present an efficient approach to the computational prediction of equilibrium isotope fractionation between 
complex minerals and fluids at high $P$ and $T$.
Both solids and fluids are treated as extended systems by application of periodic boundary conditions in all three spacial directions.
We will demonstrate that at $T \rm >600\,K$ the fractionation factor can be computed by considering the force constants acting 
on the fractionating element only. 
Both solid and fluid supercells should be big enough to avoid significant interaction between atoms and their periodic images.
In our investigation we use cells at least $5\,\rm\AA$ wide in each spacial dimension.
A representative distribution of relevant coordination environments in the fluid structure is obtained by performing
Car-Parrinello molecular dynamics simulations \citep{CPMD1}.
For the calculation of the fluid fractionation factors, several random snapshots from this simulation are chosen.
The force constants acting on the fractionating element and the resulting fractionation factors are then obtained for each configuration and
the fractionation factor for the considered element in the fluid is computed as an average over the whole set of geometries.

As a test case for our approach, we have computed the fractionation factors between Li bearing aqueous fluids and three minerals, mica, staurolite and 
spodumene. For these systems, recent experimental data are available for comparison \citep{W05,W07}.
Furthermore, lithium as one of the lightest elements with two stable isotopes
produces strong isotope signatures. It strongly fractionates into aqueous fluids during fluid-rock interaction processes
and is used as a tracer of mass transfer in the subduction cycle \citep{W05}. The two stable isotopes, $\rm ^7Li$ and $\rm ^6Li$,
have respective abundances of 92.5$\%$ and 7.5$\%$. The large mass difference of $7.016003/6.015121=16.6\%$
results in a prominent fractionation of at least a few {\permil} even at high temperatures $T\rm\sim1000\,K$. 
The experimental data on Li isotopes indicate a significant influence of the Li coordination and the Li-O bond length on the 
fractionation of Li isotopes.
The heavier isotope preferentially occupies the lower coordinated sites and phases with shorter bond distance \citep{WJ11}, 
which is expected also from the theoretical point of view \citep{SMH09}.
We will show that the application of {\it ab initio} methods to Li-bearing silicates and fluids provides unique insight into the
mechanisms driving equilibrium Li-isotope fractionation on the atomic scale.

\begin{table*}
\caption{The $\beta$ factors for mica (columns 2-5) and fractionation factors between mica and spodumene (last column) computed at $T\rm=650\,K$ for various mica polytypes and Li substitution sites.
All values are given in $\permil$. The measured value is that of \citet{W07}.
}
\label{T2}      
\centering          
\begin{tabular}{l c c c c c c}     
\hline\hline       
Mineral & Li1 & Li2 & Li3 & Average & $\rm \Delta^7Li_{mc.-spd.}$ \\
\hline                    
 1M & 13.9 & 14.9  & - & 14.6 & +4.7$\pm0.9$  \\
 occupation & 0.3 & 0.97 & -  & &   \\
 2M1 & 13.9 & 13.6 & - & 13.7 & +3.8$\pm0.9$  \\
 occupation & 0.38 & 0.92 & -  & &   \\
 2M2 & 13.8 & 13.4 & - & 13.6 & +3.7$\pm0.9$  \\
 occupation & 0.37 & 0.95 & -  & &   \\
 3T & 12.2 & 14.9 & 12.1 & 13.6 & +3.7$\pm0.9$  \\
 occupation & 0.7 & 0.89 & 0.14  & &   \\
 exp. & &  & & & +2.5$\pm$1.0 \\

\hline                  
\end{tabular}
\end{table*}

\section{Theoretical model \label{TM}}
The mass-dependent equilibrium isotope fractionation is driven by the change in the molecular and crystalline vibration frequencies
resulting from the different mass of the isotopes.
The fractionation between species and an ideal atomic gas is called the $\beta$ factor or the reduced partition function ratio (RPFR) 
and in the harmonic approximation is given by the formula:
\begin{equation} \beta=\prod_{i=1}^{N_{dof}}\frac{u_{i}^{*}}{u_{i}}\exp{\frac{(u_{i}-u_{i}^*)}{2}}\frac{1-\exp(-u_i)}{1-\exp(-u_i^*)}, \label{beta}\end{equation}
where $u={\hbar\omega_i}/{k_BT}$, $\hbar=h/2\pi$ is the reduced Planck constant, $\omega_i$ 
the vibrational frequency of the $i$-th degree of freedom,
$k_B$ is the Boltzmann constant,
$N_{dof}$ is the number of degrees of freedom, which for the $N$ being the number of atoms in 
the considered system (molecule, mineral or fluid)
is equal to $3N-5$ for a diatomic molecule, $3N-6$ for multiatomic molecules and $3N$ for crystals, 
and a star symbol marks the heavier isotope.
Despite requiring only the knowledge of the vibrational frequency spectrum, 
the above formula accounts also for the translational and rotational motions of a molecule \citep{CCH01}.
Because of the Redlich-Teller product rule, equation \ref{beta} is also valid for minerals (but with the product running 
to $3N$), if the crystal is represented as a big molecule \citep{CCH01}. 
The fractionation factor between two substances A and B, $\alpha_{A-B}$ 
is computed as the ratio of the relevant $\beta$ factors, which
for $(\beta-1)\sim 10^{-3}$ is well approximated by its differences:
\begin{equation} \alpha_{A-B}=\beta_A/\beta_B\simeq\Delta_{A-B}=\beta_A-\beta_B.\end{equation}
The calculation of the $\beta$ factor requires only the knowledge of the vibrational properties 
of the considered system computed for the two different isotopes.
However, computation of the whole vibrational spectra of complex, multiparticle 
minerals or fluids requires substantial computational resources and is currently limited to systems containing a few dozens of atoms.
Any approach that would allow for a substantial reduction of computational time and computationally efficient treatment of complex systems is highly desired.
\citet{BM47} have shown that in case of $u<2$ the isotope fractionation can be computed from the knowledge of the force constants acting on 
the atom of interest. The $\beta$ factor (Eq. \ref{beta}) can then be approximated by:
\begin{equation} 
\beta\simeq 1+\sum_{i=1}^{N_{dof}}\frac{u_i^2-u_i^{*2}}{24}=1+\frac{\Delta m}{m m^*}\frac{\hbar^2}{24 k_B^2T^2}\sum_{i=1}^3 A_i \label{BAPX} 
\end{equation}
where $A_i$ are the force constants acting on the isotopic atom in the three perpendicular spacial directions (x, y and z),
$\Delta m=m^*-m$ and $m$ is the mass of the fractionating element.
For clarity we will call the formula \ref{BAPX} {\it the single atom approximation} through the paper.
The validity criterion, $u=\hbar\omega/k_{B}T<2$, restricts the usage of the formula to frequencies $ \omega\,{\rm[cm^{-1}]}<1.39\,T\rm{[K]}$. 
As it is rare that $\omega\rm>>1000\,cm^{-1}$ 
(with the exception of the vibrations involving hydrogen), the formula is usually valid for high temperatures $T\rm>600\,K$. 
In the case of Li, $\omega\rm<600\,cm^{-1}$ and the formula is valid down to $T\rm\sim450\,K$.
This gives us the opportunity to simplify the calculations by considering the force constants acting on one atom of interest instead of all
atoms constituting the considered system. 
For large systems containing hundreds of atoms the speed up in the calculations can be significant as the full normal mode analysis 
of an $N$-atoms system requires $N$ times more computations than computing a single atom. For instance for a system containing 100 atoms the calculations
using the single atom approximation are 100 times faster.
We will show that the computation of isotope fractionation factors from the knowledge of the force constants acting upon the element of interest 
allows for efficient calculation of Li isotope fractionation between complex silicates, such as spodumene, Li-micas and Li-staurolite, and aqueous solutions.

One important aspect of the method is its usage for the calculation of isotope fractionation in crystals. 
In principle in order to compute the $\beta$ factors for crystals one has to account for dispersion.
In a solid the phonon frequencies are identified by a $q$-vector in a reciprocal-space, which requires extension of the product
in Eq. \ref{beta} beyond the number of atoms and adding multiplication over the $q$-vector grid (see Eq. 16 of \citet{ML07}).
However, considering the $\rm ^{26}Mg/^{24}Mg$ fractionation in Mg-bearing minerals \citet{SCh11} has shown recently that 
for minerals of multiatomic structure ($N>20$) considering a single phonon wave-vector 
is sufficient for getting very accurate 
$\beta$ factors even at $T=\rm 300\,K$ (error of 0.1\permil). At $T=1000\rm\,K$ the error is negligible and in the order of 0.01\permil.
This finding and the computation of $\beta$ factors considering the single atom approximation
reduce the computational load required to compute the fractionation factors to calculation of only the force constants acting upon fractionating element.
This allows for computer-aided investigation of isotope fractionation in complex minerals and fluids containing hundreds of atoms.

\begin{figure*}[t]
\includegraphics[angle=270,width=4.in]{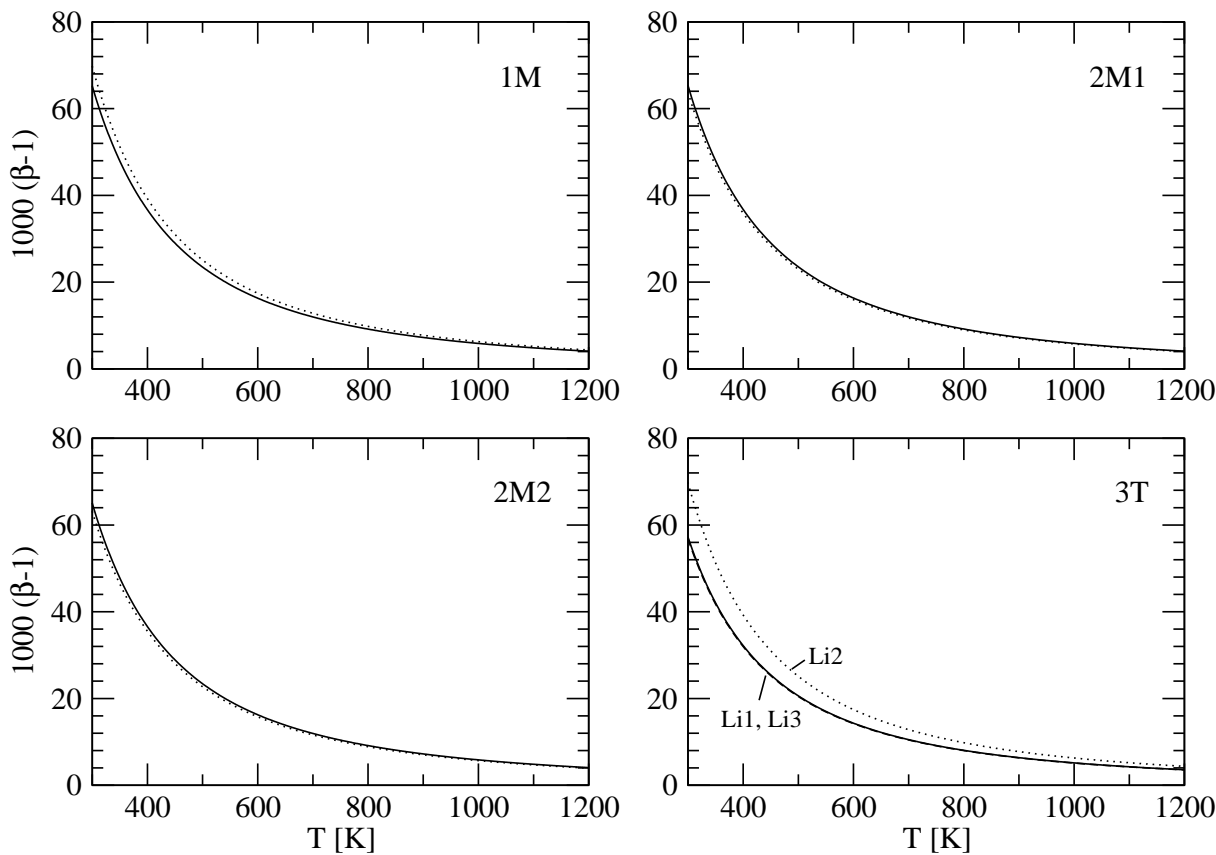}
\caption{ The $\beta$ factor for various polytypes of mica. The lines represent the results 
for isotope substituted on different Li sites: Li1 (solid lines), Li2 (dotted lines) and Li3 (dashed line). \label{FIG2}}
\end{figure*}

\section{Computational approach \label{CA}}

The calculations of $\beta$ factors of crystals and aqueous solutions were performed by applying density functional theory (DFT) methods, 
which are currently the most efficient methods allowing for treating extended many particle systems quantum-mechanically. 
For that purpose we used the CPMD code \citep{CPMD2}, which is especially suited for {\it ab initio} simulations of fluids. 
In order to reach consistency with previous work on Li-bearing aqueous fluids \citep{JW09},
we used the BLYP exchange-correlation functional \citep{BECKE88,LEE88}, a plane wave basis set and an energy cut-off of $70\,\rm Ryd$
for geometry relaxations and molecular dynamics simulations and of $140\rm\,Ryd$ for computation of vibrational frequencies and force constants.
The much higher cut-off used for derivation of the vibrational frequencies and force constants was essential to obtain the converged $\beta$
factors. Norm-conserving Goedecker pseudopotentials were applied for the description of the core electrons \citep{GOE96}. 
For both crystalline solids and aqueous solutions, periodic boundary conditions were applied. 
The solids were represented by large cells containing at least 40 atoms. The number of atoms used in the 
crystal calculations together with the lattice parameters of modeled crystals are summarized in table \ref{T1}. 
The lattice constants used in our calculations 
resemble those determined by \citet{W05,W07}. 
The atomic positions of the crystal structure were relaxed to the equilibrium positions to minimize the forces acting on the atoms.
The aqueous solution was represented 
by a periodically repeated box containing up to 64 water molecules and one Li atom. 
The Li$^+$ cation in the fluid was charge balanced by an F$^-$ anion.
The pressure and temperature conditions were chosen to be close to the experimental conditions of \citet{W05,W07}.
The pressure of aqueous solution for a given temperature and volume was calculated according to the equation of state of
\citet{WP02}. The {\it ab initio} molecular dynamics simulations (AIMD) were preformed for fixed temperature and volume 
using Car-Parrinello scheme \citep{CPMD1}. The temperature during each run was controlled by a Nos{\'e}--Hoover chain thermostat \citep{NK83,H85}.
For each $T-V$ conditions at least $10\rm\,ps$ long trajectories have been generated with an integration step of $0.12\rm\, fs$.
The sum of the force constants needed for computation of the $\beta$ factors from equation \ref{BAPX} was computed using finite displacement 
scheme 
by fixing the positions of all the atoms except the fractionating element. The full normal mode analyses
were performed using the same method, but displacing all the atoms constituting the considered system. The frequencies were obtained through the
diagonalization of the full dynamical matrix \citep{S04} as implemented in CPMD code. In case of solids the atomic structures taken for computations
of $\beta$ factors were those obtained after relaxation of atomic positions to minimize the forces for given lattice constants.
For fluids the $\beta$ factors were computed on the ionic configuration snapshots extracted uniformly in $0.1\rm\,ps$
intervals along the $10\rm\,ps$ long molecular dynamics trajectories. The calculations were performed with the positions of water molecules 
fixed to the molecular dynamics configurations and the Li cation was relaxed to the equilibrium position.
The effect of the continuous medium on the derived fractionation factors was studied by additional computations of 
$\rm Li(H_2O)_n^+$ isolated clusters.
For that purpose we used a large, isolated simulation box of the length of $16\rm\,\AA$, forcing the charge density to be zero at the boundary, 
as implemented in CPMD code. 

\begin{figure}[t]
\includegraphics[angle=270,width=3.5in]{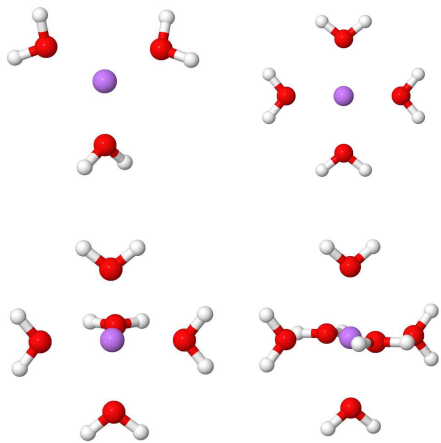}
\caption{ The structures of $\rm [Li(H_2O)_n]^+$ clusters. \label{FIG3}}
\end{figure}

The error in the computed value of the $\beta-1$ and $\Delta$ fractionation factors
we estimate from an average error of vibrational frequencies computed using chosen
DFT method. \citet{FS95,MT02} estimated the errors made in calculations of vibrational frequencies
of small molecules using different DFT functionals. 
According to these works BLYP functional systematically overestimates the frequencies by $\sim3.5\,\%$
with the deviation from the mean value of $\sim1\,\%$. 
Therefore, we expect that using BLYP functional 
the $\beta-1$ and $\Delta$ values are systematically overestimated by $\rm 7\,\%$
and that in addition there is $\rm 2\,\%$ error in derived $\beta-1$ factors.
We notice that in order to correct for the systematic errors 
some authors (for example \citet{SCh11}) scale the DFT vibrational frequencies usually by a frequency independent 
scaling factor, which could be derived from the match to the experimental measurements. We decided not to use such a scaling as we intent to
test the ability of DFT methods to predict the stable isotope fractionation factors from first principles without introducing free parameters,
or making constraints to the experimental data.

\begin{figure}[t]
\includegraphics[angle=270,width=2.5in]{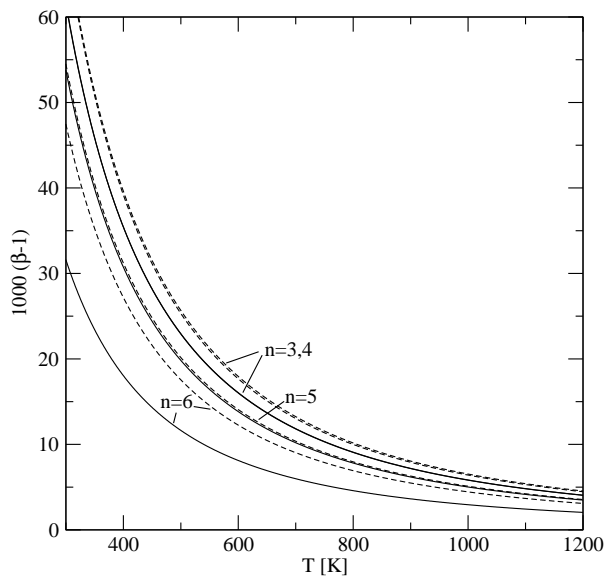}
\caption{ The $\beta$ factors for $\rm [Li(H_2O)_n]^+$ clusters. The lines represent the results 
for n=3,4,5 and 6 (from top to bottom) of this work (solid lines) and using frequencies computed in \citet{YMMW01} (dashed lines).
The results for n=3 and n=4 are nearly identical and hardly resolved in the figure.
\label{FIG4}}
\end{figure}

\section{Results and discussion}

\subsection{Solids}

\subsubsection{Representation of the silicates}

\begin{figure}[t]
\includegraphics[angle=270,width=2.5in]{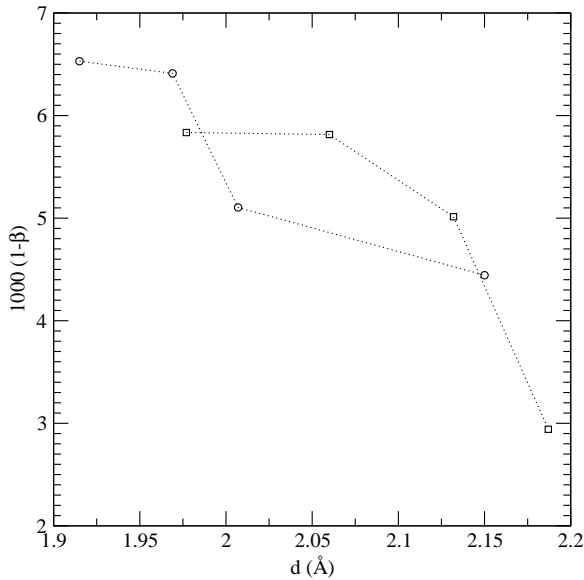}
\caption{ The dependence of the $\beta$ factors of $[\rm Li(H_2O)_n]^+$ clusters on the Li-O distance in the clusters derived using vibrational frequencies of \citet{YMMW01} (circles)
and ours (squares). The coordination runs from 3 (left) to 6 (right). The dotted lines connecting the data points are added to visualize the trend. \label{FIG5}}
\end{figure}

The lattice parameters of the modeled crystalline solids are the experimental values found in the literature.
For staurolite, the refined crystal structure of \citet{CM02} was used.
As in the experiment of \citet{W07} Mg-staurolite was used instead of Fe-staurolite, in order 
to reproduce closely the experimental conditions we replaced all the Fe atoms in the modeled structure with Mg atoms. 
In staurolite Li is a trace species. 
Following the assignment of \citet{W07} we assumed that it occupies one of the T2 sites, i.e.
the 4-fold coordinated site occupied by Mg atoms, and that there is only one substitution site. 
The constructed model contains a single unit cell of chemical composition $\rm Al_{18}^{[6]} (Li_{1} Mg_{3})^{[4]} Si_{8}^{[4]}  O_{45}(OH)_{3}$,
where in square brackets we denote the coordination number.
The chosen composition and lattice parameters closely resembles the ones determined for Mg-staurolite by \cite{W07}.

Spodumene is the simplest crystal investigated here. The modeled structure is that of \citet{CS73}. 
The chemical composition of the unit cell used in the investigation is $\rm (Li_8 Al_8)^{[6]} Si_{16}^{[4]} O_{48}$, which is exactly the chemical compositions of spodumene synthesized 
and used in the isotopic measurements by \citet{W05}.

Comparing with staurolite and spodumene the Li-bearing mica obtained in the experiments by \citet{W07} is a complex
silicate system containing
different polytypes with relative abundances varying significantly between different samples (see table 3 of \citet{W07}).
Following structure determination of \citet{W07} in our investigation we consider four mica polytypes: 1M, 2M1, 2M2 and 3T. 
The structural parameters and literature sources are given in table \ref{T1}. 
In order to model the minerals synthesized in \citet{W07} experiment we represent the different mica polytypes by the supercells of the following chemical compositions: 
$\rm K_{2} (Li_{4} Al_{2})^{[6]} Si_8^{[4]} O_{20} (OH)_{4} $ for 1M mica,  
by $\rm K_{4} (Li_{8} Al_{4})^{[6]} Si_{16}^{[4]} O_{40} (OH)_{8}$ for 2M1 and 2M2 micas, and $\rm K_{3} (Li_{6} Al_{3})^{[6]} Si_{12}^{[4]} O_{30} (OH)_{6}$ for 3T mica.

\subsubsection{Li isotope fractionation between minerals}
Staurolite and spodumene have a single Li occupation site.
In staurolite, Li substitutes for Mg in a four-fold coordinated site, while in spodumene Li occupies the six-fold coordinated 
M2 site of pyroxenes. In both cases Li is bounded to oxygen atoms only. 
The computed $\beta$ factors for both silicates are given in figure \ref{FIG1}.
We give the results of two sets of calculations:  (1) considering force constants acting upon Li atom only using Eq. \ref{BAPX} and 
(2) performing full normal mode analysis, i.e. computing full phonon spectrum at the gamma point and using Eq. \ref{beta}.
This provides an explicit test of the single atom approximation outlined in section \ref{TM}.
The $\beta$ factors derived using both methods are essentially identical and only deviate slightly at low temperatures,
which is expected. \citet{W07} and \citet{W05} measured the Li isotope fractionation between these two minerals and the aqueous solution. 
According to their measurements the fractionation between staurolite and spodumene is $2.7\pm 1.0\,\permil$ at $1200\rm\,K$  
and $3.7\pm 1.0\,\permil$ at $1000\rm\,K$. The calculated values, which are given by the differences between $\beta$ factors at the considered temperatures
are $\rm \Delta^7Li_{str.-spd.}=\beta_{str.}-\beta_{spd.}=3.7\pm0.5\,\permil$ and $4.6\pm0.5\,\permil$ respectively and therefore in good agreement with the experiment.
We will show that because of thermal expansion effect and different experimental pressures ($3.5\rm\,GPa$  with staurolite and $2.0\rm\,GPa$ in experiments with spodumene),
the computed $\rm \Delta^7Li_{str.-spd.}$ is overestimated by $1.1\,\permil$, bringing the prediction to even better agreement with the measured data.

The case of mica is more complex as it contains different polytypes and Li substitution sites. 
In the experiment of \citet{W07} the measured mica samples contained various combinations of 1M, 2M1, 2M2 and 3T polytypes.
To account for that we have computed the $\beta$ factors for all the outlined polytypes and 
Li substitutions sites. The results are given in figure \ref{FIG2}. It is clearly visible that
both the polytype and Li substitution sites impact slightly the value of computed $\beta$ factors.
This is because the different structural environments result in slightly different Li-O bond lengths, 
although the Li coordination is the same in all cases. 
The largest difference is visible in case of 3T polytype, where $\beta$ factor computed for Li2 site is higher
than for the other Li sites and polytypes. This is because even after atomic relaxation this particular site exhibits the shortest Li-O bonds
with the strongest Li-O bond shorter by $\sim0.05-0.1\,\rm \AA$ comparing to other Li sites and polytypes.
\citet{W07} showed that at approximately $T\rm=650\,K$ the fractionation between mica and spodumene minerals is $2.5\pm1\,\permil$. 
The results of our calculations for that temperature are reported in table \ref{T2}. Here we derived the $\beta$ factors for
the different mica polytypes by taking the statistical average over the $\beta$ factors computed for each Li site.
The contribution of each site is weighted according to the occupation of the particular site by Li atom, which is also given in table \ref{T2}.
The calculations predict 
the correct fractionation direction, i.e. $\rm \Delta^7Li_{\rm mica-spd.}>0$ and the experimental fractionation factor 
within uncertainties of the calculations (which we estimate at $\sim0.9\,\permil$ at considered temperature) but slightly overestimate the measured value. 
We will show later in the discussion of the fractionation between solids and fluid that accounting for thermal expansion of the crystals
the reported computed values decrease by $\sim 0.3\,\permil$ further improving the agreement with the experiment.

\subsection{Fluid}
\subsubsection{Cluster approach}

\begin{figure}[t]
\includegraphics[angle=270,width=2.5in]{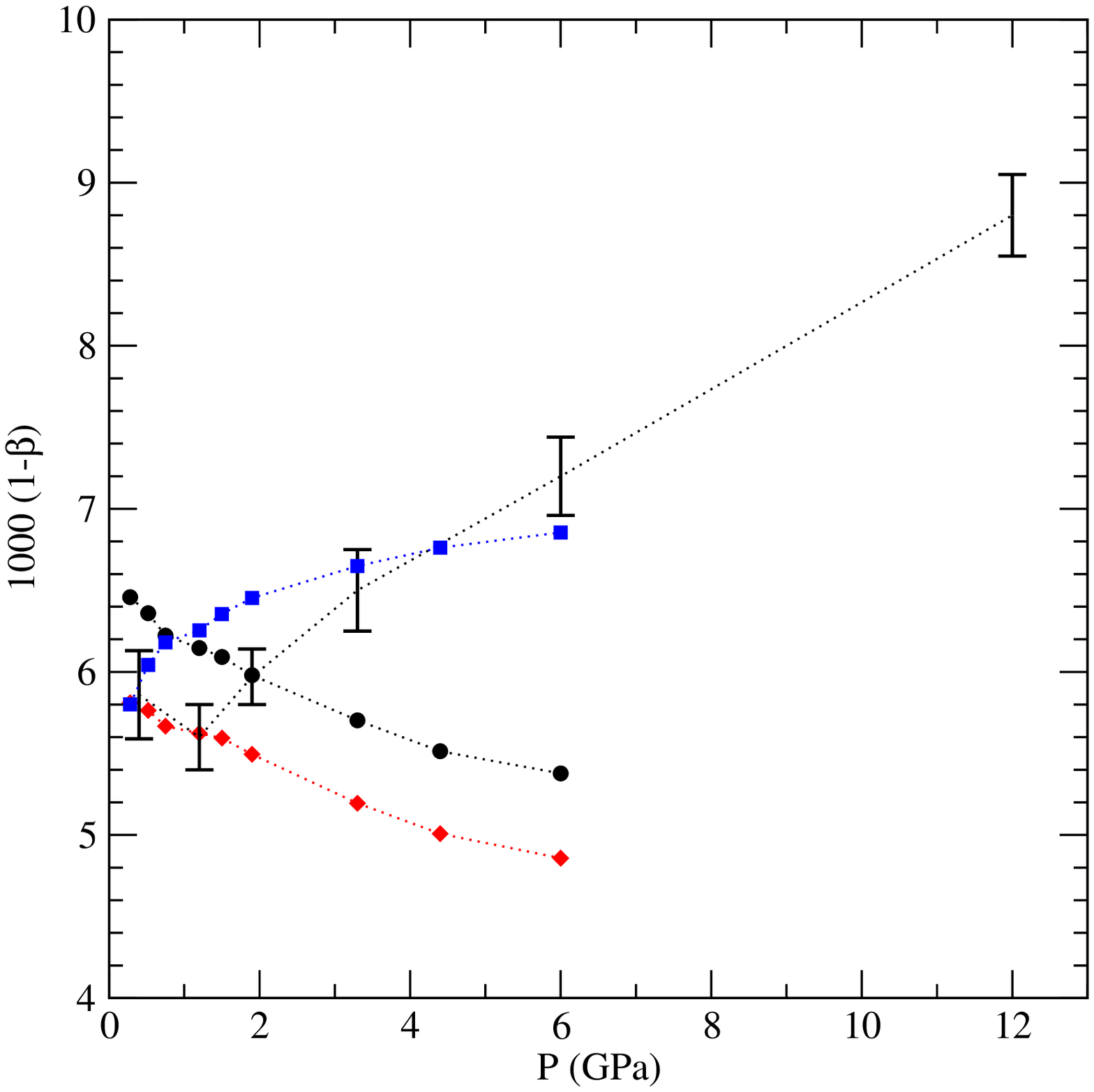}
\caption{ The pressure dependence of the $\beta$ factor for Li in the fluid computed based on the cluster approach using 
the vibrational frequencies of \citet{YMMW01} (circles), the full frequency spectrum computed for clusters in this work (diamonds) and using the \citet{BM47} approximation (their Eq. (21))
together with Li-O symmetric stretching frequencies of \citet{YMMW01} (squares).
The bars represent the $\beta$ factors computed along the {\it ab initio} molecular dynamics trajectories and their width represent the uncertainties in computed values.
The dotted lines connecting the data points are added to visualize the trend.
\label{FIG6}}
\end{figure}

\begin{table}
\caption{The distribution of coordination number of Li in aqueous solutions computed by \citet{JW09}.}
\label{T3}      
\centering          
\begin{tabular}{l c c c c}     
\hline\hline       
P[GPa] & 3  &  4 &   5 &   6 \\
\hline                    
0.28 & 0.68 &  0.30 &  0.02 &  0.00 \\ 
0.52 & 0.41 &  0.52 &  0.07 &  0.00 \\
0.75 & 0.31 &  0.53 &  0.15 &  0.01 \\
1.2 & 0.25 &  0.54 &  0.21 &  0.01 \\  
1.5 & 0.15 &  0.60 &  0.24 &  0.01 \\ 
1.9 & 0.10 &  0.58 &  0.29 &  0.03 \\  
3.3 & 0.04 &  0.47 &  0.38 &  0.11 \\  
4.4 & 0.03 &  0.35 &  0.45 &  0.17 \\  
6.0 & 0.02 &  0.29 &  0.49 &  0.20 \\  

\hline                  
\end{tabular}
\end{table}

In most of the recent work on {\it ab initio} computation of the stable isotope fractionation
in aqueous solutions the isolated cluster approach is used in which a considered species is surrounded 
by the hydration shell and the whole structure is optimized assuming $T\rm=0\,K$ \citep{YMMW01,DK08,HS08,SMH09,Z09,HS10,RC10,RB10,Z09,Z10}. 
However, at high temperatures and pressures the hydration
shell surrounding lithium ion is not static but exhibits strong dynamical character \citep{JW09} and compression
impacts its structure \citep{WJ11}.
The important questions are on the impact of these effects on the equilibrium isotope fractionation and how well these effects
can be described with the widely used cluster approach.
In order to address these problems
we performed set of calculations involving $\rm [Li(H_2O)_n]^+$ clusters.
The clusters used in the investigation are illustrated in figure \ref{FIG3}. 
Following the work of \citet{YMMW01} we computed the $\beta$ factors for isolated $\rm [Li(H_2O)_n]^+$ clusters for $\rm n=3,4,5,6$, relaxing
the structures to equilibrium positions and computing the full frequency spectra. 
The spectra were then used to compute $\beta$
factors according to Eq. \ref{beta}. In the same way we also computed the $\beta$ factors
using frequencies derived by \citet{YMMW01} obtained with the restricted Hartree-Fock method (RHF). Both results are given in figure \ref{FIG4}. 
The $\beta$ factors computed with the frequencies of \citet{YMMW01} are higher than the ones derived with DFT frequencies except in the $n\rm=5$ case,
for which both calculations predict the same values. This may be related to different cluster structures used in the calculations
as positions of hydrogen atoms are not provided in details by \citet{YMMW01}. 
An interesting observation is illustrated in figure \ref{FIG5},
where the $\beta$ factor is plotted as a function of Li-O bond length. With increasing $n$ the Li-O bond length increases, as the water shell 
containing more water molecules has to relax outwards creating more space for additional molecules. 
The increase in the bond length results in a decrease of the $\beta$ factor.
This has an important implication on the pressure dependence of the $\beta$ factors derived using the cluster approach.

\begin{figure}[t]
\includegraphics[angle=270,width=2.5in]{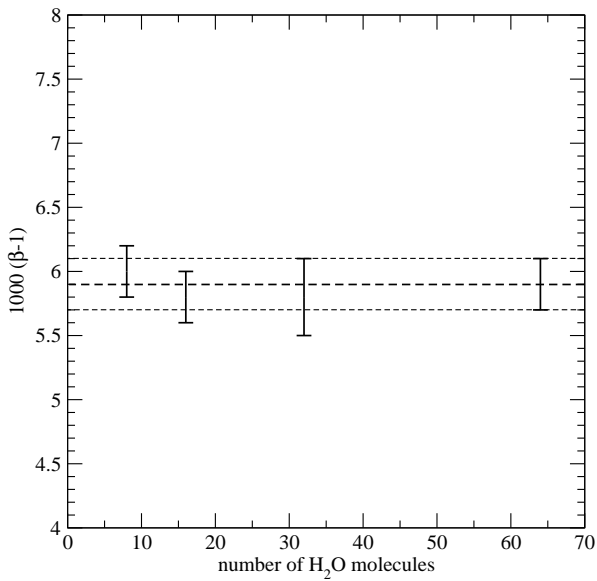}
\caption{ The dependence of the 
$\beta$ factor of fluid on the size of the simulation cell. The thick and thin dashed lines 
represent the average value and the uncertainty limits of $\beta$ factor computed on system
containing $62\rm \, H_2O$ molecules. \label{FIG7}}
\end{figure}

\begin{figure}[t]
\includegraphics[angle=270,width=2.5in]{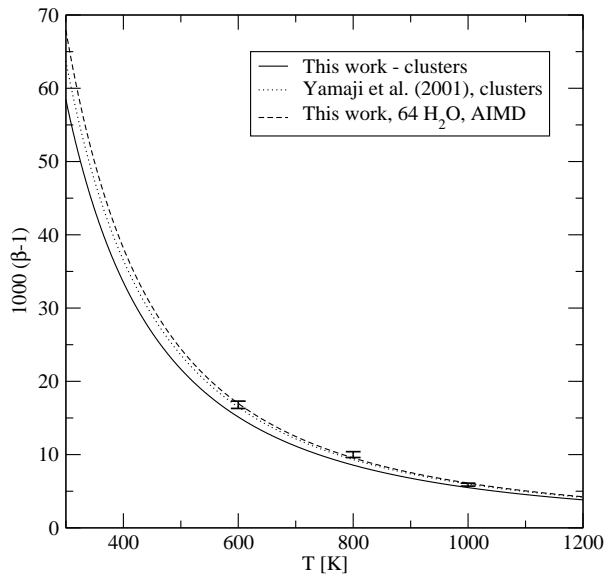}
\caption{ The $\beta$ factors for Li$^+$ aqueous solution at $1.9\rm\,GPa$ obtained by 
$\rm[Li(H_2O)_n]^+$ cluster approach and AIMD simulations. The lines represent the results of this work (solid line) 
and the $\beta$ factors obtained using frequencies of \citet{YMMW01} (dotted line). The dashed line represents the fit
to the $\beta$ factors given by bars, which indicate the uncertainty in the computed values, and computed at different 
temperatures as an average over the AIMD trajectories.
\label{FIG8}}
\end{figure}

Having both results for clusters we attempted to investigate the pressure effects on the $\beta$ factors.
We do that by averaging the $\beta$ factors over the statistical distribution of $\rm [Li(H_2O)_n]^+$
complexes in aqueous solution, which is pressure dependent. \citet{JW09} have shown that in the pressure range
from $1$ to $6\rm\,GPa$, the Li coordination by oxygens increases smoothly from preferentially four-fold to five-fold coordination.
At $2\rm\,GPa$, which corresponds to the experimental conditions of \citet{W05,W07}, the mean Li coordination is about 4.2.
We took the probability distribution of \citet{JW09}, which is given in table \ref{T3}, and derived the pressure-dependent $\beta$ factors 
as a statistically weighted average of the $\beta$ factors derived for $\rm [Li(H_2O)_n]^+$ clusters. The results are given in figure \ref{FIG6}. 
Both results derived on the two $\beta$ factor estimations predict decrease of the Li isotope fractionation with increase in pressure.
This is because at higher pressure the more coordinated and with longer Li-O bond lengths $\rm [Li(H_2O)_n]^+$ structures are preferred,
which results in lower $\beta$ factors.
This finding is counter intuitive, as one should expect that the compression of the fluid should lead to the shortening of the
Li-O bonds, elevated vibrations and resulting higher $\beta$ factors. In figure \ref{FIG6} we also give the estimation 
of $\beta$ factors computed from the knowledge of the Li-O totally symmetric stretching frequencies using rough approximation of \citet{BM47} 
(their Eq. 21) with the relevant frequencies of \citet{YMMW01} and the $\rm [Li(H_2O)_n]^+$ clusters probability distribution of \citet{JW09}.
In the \citet{BM47} approximation the $\beta$ factor is proportional to the square of the totally symmetric stretching frequency, $\nu_s$, and
the cluster size, i.e. $\beta\sim\nu_s^2n$. As with increasing the cluster size, $\nu_s$
decreases by $\sim20\%$, the largest effect on the isotope fractionation computed using the \citet{BM47} method comes from the coordination (cluster size).
The resulting pressure-dependent $\beta$ factor shows the desired tendency.
It increases with the size of the cluster, which causes the increase in pressure as is seen in figure 
\ref{FIG6}. We will show that the simulation of continuous media is required
for proper investigation of the effect of the compression and to obtain realistic isotope fractionation signature of high $P$ fluids.

\begin{figure}[t]
\includegraphics[angle=270,width=2.5in]{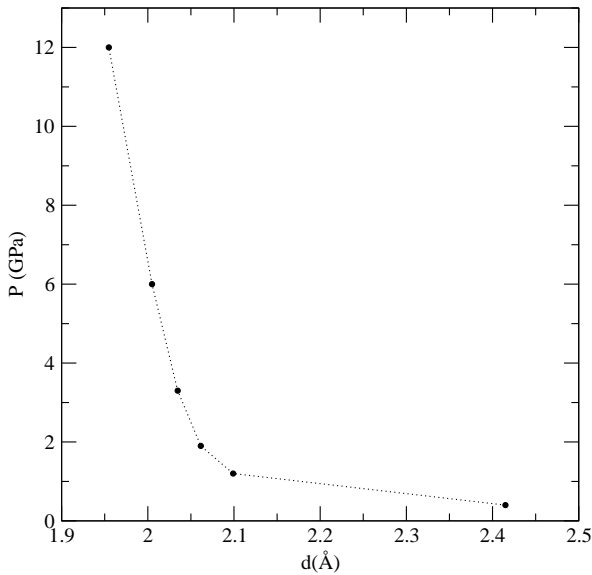}
\caption{The average Li-O distance between Li and the three closest O atoms in aqueous fluid as a function of pressure. \label{FIG6a}}
\end{figure}

\subsubsection{Molecular dynamics approach}
In order to fully account for the pressure effects, spacial continuity of the fluid and its dynamical motion we produced $10\rm\,ps$ 
long molecular dynamics trajectories of systems consisting of 64 H$_2$O molecules and one Li ion for different $T=1000\rm\,K$, $800\rm\,K$ and $600\rm\,K$ and
pressure of $1.9\rm\,GPa$, which closely resembles the experimental conditions of \citet{W05,W07}. The corresponding simulation box length is 
$12.17\rm\,\AA$ at $T=1000\rm\,K$. We note that the thermal effects on the pressure 
will require to use a supercell of $\sim1\,\%$ smaller box length for $T\rm=600\,K$, a small effect which we omitted.
As the oxidation state of Li in the aqueous solution is $+1$, following \citet{JW09} we added a F atom to the system as a charge compensator.
An interesting question is on the impact of the system size on the computed $\beta$ factors.
In order to investigate this problem we computed $10\rm \,ps$ length trajectories also for simulation boxes containing 8, 16 and 32 H$\rm_2$O molecules
for $T=1000\rm\,K$ and pressure of $1.9\rm\,GPa$. The resulting $\beta$ factors are given in figure \ref{FIG7}. Within the accuracy of the
calculation the $\beta$ factor is system size independent and in principle small systems containing 8 $\rm H_2O$ atoms
can be used in the investigation. This substantially reduces the required computational time. As the current implementations of 
plane-wave DFT methods scale
as $N^2-N^3$, with $N$ being the number of particles in the system (number of electrons), the computation time gained reducing the number of particles in the computational 
box could be significant. 
In our calculations switching from a system containing 64 water molecules to 8 the gain is a factor of $\sim$85. 
Nevertheless for our calculations we used the simulation box containing 64 water molecules.
In order to obtain the temperature dependent $\beta$ factor at $P\rm=1.9\,GPa$ we fitted by the least squares procedure the formula
$\beta=1+A/T^2$ to the $\beta$ factors calculated at the three temperatures. The parameter of the fit is $A=6.112\cdot10^{-3}$ for temperature expressed in units of $\rm10^3\,K$.
The resulting $\beta$ factor 
as a function of temperature at $P\rm=1.9\,GPa$ 
is given in figure \ref{FIG8} together with the already discussed predictions using the cluster approach.
Interestingly, the molecular dynamics $\beta$ factor is in good agreement with the value obtained by using clusters approach with \citet{YMMW01} frequencies. 
The difference between our cluster and MD calculations is also moderate, $0.6\,\permil$ at $1000\rm\, K$ and $1\,\permil$ at $800\rm\, K$.
However, the agreement between both types of calculations is only reached at lower pressures ($P\rm<2\,GPa$), which will be discussed in the next paragraph.
In order to check the validity of the single atom approximation outlined in section \ref{TM} for fluids we computed $\beta$ factors
with the full frequency spectra obtained for selected configurations. We found only negligible deviation of the resulted $\beta$ factors from the ones 
derived considering force constants acting on the Li atom only.

\begin{figure*}[t]
\includegraphics[angle=270,width=6.in]{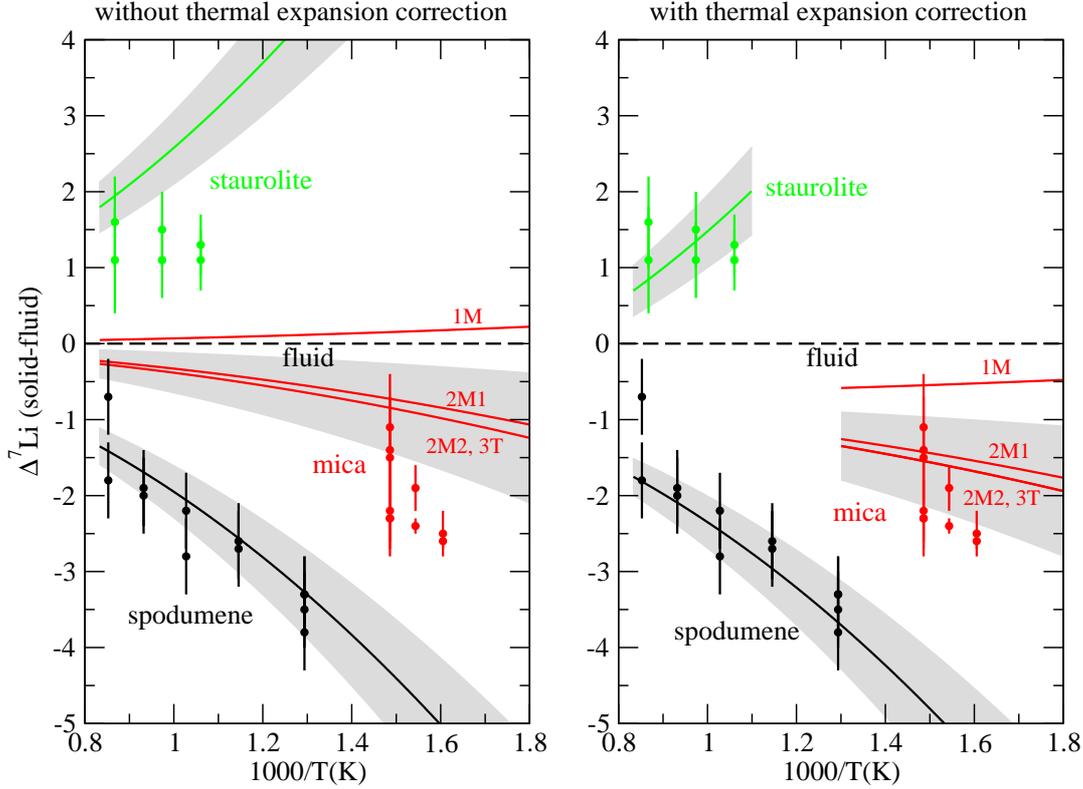}
\caption{The solid-fluid isotope fractionation between Li-bearing minerals and aqueous solution. 
The lines represent the computed values and the shadowed areas reflect the uncertainties computed assuming the computational error evaluation described in section \ref{CA}.
1M, 2M1, 2M2 and 3T lines indicate results obtained for different mica polytypes.
The points are the measured values of \citet{W05,W07}.
Left and right panels represent the result without and with thermal expansion correction discussed in the text.
The thermal expansion correction is made by a constant negative shifts of the solid-fluid fractionation factors of $-0.6\rm\,\permil$ for staurolite,
$-0.7\rm\,\permil$ for mica and $-0.4\rm\,\permil$ for spodumene. The pressure assumed for the computations is $\rm 2\,GPa$ \label{FIG9}. 
The computed values for staurolite given in right panel are shifted by additional $-0.5\rm \,\permil$, the correction due to 
the higher experimental pressure $P\rm=3.5\,GPa$, as discussed in the text.}
\end{figure*}

In the $\rm [Li(H_2O)_n]^+$ cluster calculations
we obtained an unexpected result indicating that $\beta$ factor should decrease with pressure,
which we found counter intuitive. In order to investigate the pressure impact on the $\beta$ factor accounting for the continuity on the medium and
its pressure-driven compression we computed the fractionation factors at $T\rm=1000\,K$ and different pressures on a system containing $8\rm\,H_2O$ molecules.
In each case $10\rm\,ps$ long trajectories were generated and $\beta$ factors were computed on a set of atomic configurations extracted uniformly along the trajectories. 
The result is given in figure \ref{FIG6}. We clearly see that for pressures $P\rm>2\,GPa$, as the effect of compression, the $\beta$ factor increases monotonically with 
increasing pressure.
The reason for that is the small decrease in the mean Li-O bond length (measured as the average distance between Li and the three closest neighbors) 
with increasing $P$, which is opposite to the result using clusters approach, and the coordination, as shown in figure \ref{FIG6a}.
This finding is in line with results of \citet{WJ11}, who found that the mean Li-O distance increases for the pressures up to $1\rm\,GPa$
and remains constant at higher pressures. This explains why the computations using cluster approach, in which the increase of the Li-O bond lengths 
with the increase in the cluster size, and therefore pressure, is also observed, produce good pressure dependence of $\beta$ factor 
at low pressures, as illustrated in figure \ref{FIG6}. On the other hand this clearly shows that an isolated cluster 
is not a good representation of high $P$ and $T$ fluid and can not be used for the computation of the $\beta$ factors in fluids at extreme conditions. 
Continuity and compressibility of the fluid have to be considered in order 
to obtain realistic results. 

Although most of the experimental results to which we refer in this paper were performed at lower pressures ($2-3.5\rm\,GPa$), at which 
our results indicate small pressure effects on the fractionation (see Fig. \ref{FIG6}), \citet{WJ11}
report a measurement of Li isotope fractionation between spodumene and aqueous fluid at $T\sim900\rm\,K$ and $P\rm=8\,GPa$ to be $+0.75\pm0.5\,\permil$ 
lower than the values measured at the same temperature but lower pressures for the same systems in \citet{W05}. 
In order to check if we are able to reproduce this behavior with our method we computed the $\beta$ factor of spodumene at high $P=8\,\rm GPa$ 
by using the lattice constants of compressed spodumene determined by \citet{AA00}. Because of the high compression, the resulting $\beta$ factor 
is $3\rm\, \permil$ higher than the one derived for uncompressed solid. The $\beta$ factor of fluid at the same $(P,T)$ conditions increases by $1.9\rm\, \permil$.
This results in pressure-driven decrease of the spodumene-fluid fractionation factor ($\rm \Delta^7Li_{spd.-fluid}$) by $1.1\rm\, \permil$,
which is in good agreement with the result of \citet{WJ11}.

\subsection{Fluid-mineral fractionation}

The different experiments on Li isotope fractionation between Li-bearing minerals and aqueous solution 
at high $P$ and $T$ \citep{W05,W07} show the strongest enrichment in $^7$Li for staurolite and subsequently lighter isotopic
signatures for the fluid, mica and spodumene.
An important test for our proposed computational method is to reproduce the sequence of experimentally observed fractionation factors.
The crystal structures used in the calculation and the procedure used to compute the $\beta$ factors are described in previous sections, and the
relevant $\beta$ factors were already discussed. The computed fluid-mineral fractionation factors, $\rm \Delta^7Li_{mineral-fluid}$,
between staurolite, spodumene and mica, and aqueous solution are given in figure \ref{FIG9} 
together with the experimental values of  \citet{W07} for mica and staurolite and \citet{W05} for spodumene respectively.
The errors of the computed fractionation factors are given in the figure caption and are derived assuming uncertainty in the
computed vibrational frequencies coming from using BLYP functional, which is discussed in section \ref{CA}.
The computed curves correctly predict the fractionation sequence.
The heavy Li isotope preferentially fractionates into staurolite with respect to aqueous solution, whereas
spodumene is enriched in $^6$Li.
The computations also reproduce the experimental results for both minerals on the quantitative level within $1-1.5\,\permil$, taking into account
the uncertainties in the calculated fractionation factors, and our prediction for spodumene is ideal.
In case of mica the picture is more complicated as it has four polytypes and more than one Li substitution site. 
In figure \ref{FIG9} we plotted the average mica-fluid fractionation factors computed for different polytypes.
The resulted solid-fluid fractionation is higher than the experimental values by $\sim1-2\,\permil$, depending on the polytype.
Nevertheless, our results confirm that among the considered minerals, the fractionation between mica and the fluid is the smallest
and that on average the mica containing mixture of different polytypes should be slightly enriched with light isotope comparing with fluid.
We note that as the measurements for mica were performed at lower temperature of $\sim650\rm\,K$, the error in the calculated fractionation factor between mica and fluid 
is significant and $\sim0.6\rm\,\permil$. The straightforward comparison of our results for mica crystalline solid with the experimental data is also complicated 
as different reported measured samples of \citet{W07} contained different relative abundances of different polytypes.
We also found that of all the crystalline solids considered here mica is the most sensitive to the change in the lattice parameters
and computational setup. For instance, while the $\beta$ factors for other minerals and the fluid are well converged (within $0.1\rm\,\permil$) using the force constants obtained 
with the plane wave energy cutoff of $100\rm\,Ryd$, the resulted $\beta$ factors for mica with this setup are overestimated by $\sim1.5\rm \,\permil$
and the converged values were obtained by applying much higher cutoff of $140\rm\,Ryd$.

We notice that for staurolite and mica the solid-fluid fractionation factors are overestimated by $\sim0.5-1.5\rm\,\permil$.
However, the lattice parameters used in the calculations of the crystalline solids are the one measured at ambient conditions.
At high temperatures solids undergo thermal expansion, which should result in the lowering of
the $\beta$ factors. Observing the deviation of the computed solid-fluid fractionation factors for staurolite and mica 
we attempted to check for the effect of the thermal expansion of the lattice parameters of modeled solids
on the derived fractionation factors. According to the crystal structure data of \citet{CS73} the lattice constants of spodumene expand by $\sim\rm0.5\,\%$
at $T\sim1000\rm\,K$. Having such a pronounced effect, we recalculated the $\beta$ factors of spodumene at $T=573\rm\,K,\,723\rm\,K\, and\,1033\rm\,K$
using temperature-dependent lattice parameters of \citet{CS73}. We found that the thermal expansion of spodumene
results in $\sim0.4\rm\,\permil$ decrease in the $\beta$ factors for all the considered temperatures. Similar reduction is observed for micas.
\citet{RG99} showed that for the phlogopite 1M mica the lattice parameters increase by $\sim0.5\,\%$ at $T\rm=650\,K$.
Assuming that Li-bearing micas undergo similar expansion we computed the $\beta$ factors with the lattice parameters rescaled by $+0.5\,\%$.
The resulted $\beta$ factors are $\sim0.7\rm\,\permil$ smaller, which indicates that inclusion of thermal expansion effect lowers the 
computed mica-fluid fractionation curves by $0.7\rm\,\permil$. We also computed the $\beta$ factor for staurolite assuming the $\sim0.5\,\%$
increase of its lattice parameters at $1000\rm\,K$ \citep{HP11}. The resulted $\beta$ factor decreases by $0.6\rm\,\permil$.
In addition we notice that the measurements for staurolite were performed at higher pressure $P\rm=3.5\,GPa$ \citep{W07}.
At such elevated pressure the $\beta$ factor for fluid increases by $\sim\rm0.5\,\permil$  (Fig. \ref{FIG6}) leading to the further decrease
of the staurolite-fluid fractionation factor by $\rm0.5\,\permil$. 
The solid-fluid isotope fractionation factors resulted by applying the derived shifts in $\beta$ factors
are given in the right panel of figure \ref{FIG9}. It is clearly visible that the corrections due to the thermal expansion of crystalline solids 
and the high pressure in case of staurolite make the prediction more consistent with the measurements. 
We note that on the right panel we plotted the solid-fluid fractionation curves only 
at the temperature range corresponding to the experiment as being interested in direct comparison of the computed values with the experiments 
we applied the constant thermal expansion and pressure correction derived only at these temperatures. The respective corrections for staurolite and mica 
at other temperatures may be different.

Beside the thermal expansion effect and the uncertainties and systematic errors resulting from choice of the DFT functional
there are other effects that could potentially increase the uncertainties in the calculated fractionation factors.
These additional effects could arise from the usage of the experimental equation of state for fluid and lattice parameters 
for crystalline solids, and uncertainties in the crystalline lattice site occupations as in the case of micas.
On the other hand, it can not be guaranteed with full confidence that the experimental measurements of \citet{W05,W07}, which indicate complete isotopic exchange, 
reflect an equilibrium fractionation (see for instance \citet{LJ11})
\footnote{Although the arguments supporting the equilibrium fractionation as given in \citet{W05,W07} are convincing
and the good agreement between our predictions and the measurements corroborate that scenario.}.
Nevertheless, the good agreement between theoretical prediction and experimental data on the Li isotope fractionation between complex Li-bearing minerals
and aqueous fluid shows that the outlined method for computing the isotope fractionation of fluids and crystals is a powerful tool,
which can be successfully applied for prediction of isotopic signatures of complex Earth materials under extreme conditions.

\section{Conclusions}

We propose a computationally efficient approach for computation of the $\beta$ and isotope fractionation
factors for complex minerals and fluids at high temperatures and pressures.
We demonstrated that in order to derive the reliable $\beta$ factors for either minerals or fluids at high $T$ and $P$ 
it is sufficient to know the force constants acting on the
substituted isotope. This reduces significantly the computational time and allows for computations 
of isotope fractionation in complex materials containing even hundreds of atoms. 
In case of fluids we show that the widely used technique of representing aqueous solution as an ion-hydration-shell 
cluster is not sufficient to reproduce the isotope fractionation in aqueous solutions at elevated temperatures and pressures,
when the dynamical character of the hydration shell and the compression of the fluid have to be accounted for. 
This can be achieved by {\it ab initio} molecular dynamics simulation technique, which allows for direct access
to the dynamical distributions of water (fluid) molecules around the considered ion and proper consideration of compression effects.
The relevant isotope fractionation factors can be computed on a set of uncorrelated snapshot configurations extracted
from the molecular dynamics trajectory.

We show that in the case of Li in aqueous solution it is sufficient to compute the $\beta$ factors from the molecular dynamics simulations 
performed with a simulation cell containing a small number of atoms, which further reduces the 
computational time needed to perform the task.
A system containing a single Li, a charge compensating anion 
and 
$8\rm\,H_2O$ molecules was sufficient to obtain the accurate $\beta$ factors within the uncertainties of the
{\it ab initio} method used in the calculations.

We verify our approach by computing the Li isotopes fractionation factors between Li-bearing minerals and aqueous solutions
and their comparison with the experimental data. The computed Li fractionation factors between staurolite, spodumene, mica 
and aqueous solutions reproduce the experimental results on quantitative and qualitative levels. We show that {\it ab initio} calculations 
are able to predict the correct sequence of isotopes fractionation between considered materials as observed in the experiment.
The computed fractionation factors are within $1\,\permil$ in agreement with the measured values.
We also found that the thermal expansion of the solids affects the isotope fractionation process and its inclusion improves the agreement with the experimental data.

Our study shows that {\it ab initio} computer simulations represent a powerful tool for prediction and understanding of equilibrium stable isotope fractionation
processes between various phases including aqueous solutions at high pressures and temperatures.
We expect that with the increasing power of computers and performance of the computational software 
these methods will be extensively applied to complement analytical techniques and to interpret measured isotopic signatures.

\section*{Acknowledgements}
The authors wish to acknowledge financial support in the framework of DFG 
project no. JA 1469/4-1. Part of the calculations were performed on the IBM 
BlueGene/P JUGENE of the John von Neumann Institute for Computing (NIC).
We are also grateful to Bernd Wunder for fruitful discussions and
the associate editor Clark M. Johnson and anonymous referees for constructive comments that helped
improving the manuscript.





\bibliographystyle{model1-num-names}
\bibliography{<your-bib-database>}



\end{document}